\documentclass[preprint,12pt]{elsarticle}

\usepackage{amsthm} % theorems AMS style
\usepackage{amssymb} % math symbols
\usepackage{amstext} % typeset text in math environments
\usepackage{amsmath} % AMS math­e­mat­i­cal fa­cil­i­ties
\usepackage{bm} % bold symbols in math
\usepackage{graphicx} % enhanced support for graphics
\usepackage{epstopdf} % con­vert EPS to 'en­cap­su­lated' PDF us­ing ghostscript
\usepackage{float}
\usepackage{tikz}
\usepackage{lineno,hyperref}
\usepackage[capitalize]{cleveref} % in­tel­li­gent cross-ref­er­enc­ing
\usepackage[makeroom]{cancel} %cancel terms
\usepackage{mathtools}
\usepackage{multirow} 
\usepackage[pdf]{pstricks}
\usepackage{booktabs} 
\usepackage{lscape}
\usepackage{subfig}

\usepackage{tikz}
\usetikzlibrary{shapes,arrows}

\definecolor{red}{gray}{0.7}
\definecolor{green}{gray}{1.0}
\definecolor{blue}{gray}{0.1}
\tikzstyle{decision} = [diamond, draw, fill=blue!20, 
    text width=7em, text badly centered, node distance=3cm, inner sep=0pt]
\tikzstyle{block} = [rectangle, draw, fill=green!20, 
    text width=5em, text centered, rounded corners, minimum height=4em]
\tikzstyle{HO} = [rectangle, draw, fill=blue!20, 
    text width=5em, text centered, rounded corners, minimum height=4em]
\tikzstyle{LO} = [rectangle, draw, fill=red!20, 
    text width=5em, text centered, rounded corners, minimum height=4em]
\tikzstyle{line} = [draw, -latex']
\tikzstyle{cloud_HO} = [draw, ellipse,fill=blue!20, node distance=3cm,
    minimum height=3em]
\tikzstyle{cloud_LO} = [draw, ellipse,fill=red!20, node distance=3cm,
    minimum height=3em]

\modulolinenumbers[5]

\crefalias{subequation}{equation} % subequations counter = equations counter
\crefalias{eqnarray}{equation} % subequations counter = equations counter
\crefformat{pluraleq}{Eqs.~(#2#1#3)} % defining "plural" equations  
\Crefformat{pluraleq}{Equations~(#2#1#3)} % defining "plural" equations  

\def\bal#1\nal{\begin{align}#1\end{align}}
\def\bala#1\nala{\begin{align*}#1\end{align*}}
\def\bsub#1\nsub{\begin{subequations}#1\end{subequations}}

\journal{ArXiV}

 \biboptions{sort&compress}
\begin{document}

\begin{frontmatter}

\title{Nonlinear Fokker-Planck Acceleration for Forward-Peaked Transport Problems in Slab Geometry}

\author[osu]{J.J.~Kuczek\corref{cor1}}
\author[osu]{J.K.~Patel\fnref{patel}}
\author[osu]{R.~Vasques\fnref{vasques}}

\address[osu]{The Ohio State University, Department of Mechanical and Aerospace Engineering\\ 201 W. 19th Avenue, Columbus, OH 43210}

\cortext[cor1]{Corresponding author: kuczek.6@osu.edu; Tel: (330) 518-0333\\
Postal address: The Ohio State University, Department of Mechanical and Aerospace Engineering, 201 W. 19th Avenue, Columbus, OH 43210}
\fntext[patel]{patel.3545@osu.edu}
\fntext[vasques]{richard.vasques@fulbrightmail.org}

\begin{abstract}

This paper introduces a nonlinear acceleration technique that accelerates the convergence of solution of transport problems with highly forward-peaked scattering.
The technique is similar to a conventional high-order/low-order (HOLO) acceleration scheme.
The Fokker-Planck equation, which is an asymptotic limit of the transport equation in highly forward-peaked settings, is modified and used for acceleration; this modified equation preserves the angular flux and moments of the (high order) transport equation.
We present numerical results using the Screened Rutherford, Exponential, and Henyey-Greenstein scattering kernels and compare them to established acceleration methods such as diffusion synthetic acceleration (DSA).
We observe three to four orders of magnitude speed-up in wall-clock time compared to DSA.
 
\end{abstract}

\end{frontmatter}

\setcounter{section}{0}
\setcounter{equation}{0} 

\section{Introduction}\label{sec1}
\setcounter{equation}{0} 

Transport problems with highly forward-peaked scattering are prevalent in a variety of areas, including astrophysics, medical physics, and plasma physics \cite{HGK,aristova,multiphysics}.
For these problems, solutions of the transport equation converge slowly when using conventional methods such as source iteration (SI) \cite{adamslarsen} and the generalized minimal residual method (GMRES) \cite{gmres}.
Moreover, diffusion-based acceleration techniques like diffusion synthetic acceleration (DSA) \cite{alcouffe} and nonlinear diffusion acceleration (NDA) \cite{smithetall} are generally inefficient when tackling these problems, as they only accelerate up to the first moment of the angular flux \cite{JapanFPSA}.
In fact, higher-order moments carry important information in problems with highly forward-peaked scattering and can be used to further accelerate convergence \cite{japanDiss}.

This paper focuses on solution methods for the monoenergetic, steady-state transport equation in homogeneous slab geometry.
Under these conditions, the transport equation is given by
\begin{subequations}\label[pluraleq]{eq1}
\begin{equation}
\label{t1}
\mu\frac{\partial}{\partial x} \psi(x,\mu) + \sigma_t \psi(x,\mu) = \int_{-1}^{1} d\mu' \sigma_s(\mu,\mu') \psi(x,\mu') + Q(x, \mu), \,\,\, x\in [0, X],-1\leq\mu\leq 1  ,\\
\end{equation}
with boundary conditions
\begin{align}
\label{t2}
\psi(0,\mu) &= \psi_L(\mu), \quad \mu > 0,\\
\label{t3}
\psi(X,\mu) &= \psi_R(\mu), \quad \mu < 0.
\end{align}
\end{subequations}
Here, $\psi(x,\mu)$ represents the angular flux at position $x$ and direction $\mu$, $\sigma_t$ is the macroscopic total cross section, $\sigma_s(\mu,\mu')$ is the differential scattering cross section, and $Q$ is an internal source.

New innovations have paved the way to better solve this equation in systems with highly forward-peaked scattering.
For instance, work has been done on modified $P_L$ equations and modified scattering cross section moments to accelerate convergence of anisotropic neutron transport problems \cite{khattab}.
In order to speed up the convergence of radiative transfer in clouds, a quasi-diffusion method has been developed \cite{aristova}.
In addition, the DSA-multigrid method was developed to solve problems in electron transport more efficiently \cite{trucksin}.

One of the most recent convergence methods developed is Fokker-Planck Synthetic Acceleration (FPSA) \cite{JapanFPSA,japanDiss}.
FPSA accelerates up to $N$ moments of the angular flux and has shown significant improvement in the convergence rate for the types of problems described above.
The method returns a speed-up of several orders of magnitude with respect to wall-clock time when compared to DSA  \cite{JapanFPSA}.

In this paper, we introduce a new acceleration technique, called \textit{Nonlinear Fokker-Planck Acceleration} (NFPA).
This  method  returns  a  modified  Fokker-Planck (FP) equation  that  preserves  the  angular moments of the flux given by the transport  equation.
This preservation of moments is particularly appealing for applications to multiphysics problems \cite{multiphysics}, in which the coupling between the transport physics and the other physics can be done through the (lower-order) FP equation.
To our knowledge, this is the first implementation of a numerical method that returns a Fokker-Planck-like equation that is discretely consistent with the linear Boltzmann equation.

This paper is organized as follows.
\Cref{sec2} starts with a brief description of FPSA.
Then, we derive the NFPA scheme.
In \cref{sec3}, we discuss the discretization schemes used in this work and present numerical results.
These are compared against standard acceleration techniques.
We conclude with a discussion in \cref{sec4}.

\section{Fokker-Planck Acceleration}\label{sec2}
\setcounter{equation}{0} 
In this section we briefly outline the theory behind FPSA, describe NFPA for monoenergetic, steady-state transport problems in slab geometry, and present the numerical methodology behind NFPA.
The theory given here can be easily extended to higher-dimensional problems.
Moreover, extending the method to energy-dependence shall not lead to significant additional theoretical difficulties.

To solve the transport problem given by \cref{eq1} we approximate the in-scattering term in \cref{t1} with a Legendre moment expansion:
\begin{equation}
\label{transport1}
\mu\frac{\partial}{\partial x} \psi(x,\mu) + \sigma_t \psi(x,\mu) = \sum_{l=0}^L \frac{(2l+1)}{2} P_l(\mu) \sigma_{s,l} \phi_l(x) + Q(x, \mu),
\end{equation}
with 
\begin{equation}
\label{transport2}
\phi_l(x) =  \int_{-1}^{1} d\mu P_l(\mu) \psi(x,\mu).
\end{equation}
Here, $\phi_l$ is the $l^{th}$ Legendre moment of the angular flux, $ \sigma_{s,l}$ is the $l^{th}$ Legendre coefficient of the differential scattering cross section,  and $P_l$ is the $l^{th}$-order Legendre polynomial.
For simplicity, we will drop the notation $(x,\mu)$ in the remainder of this section.

The solution to \cref{transport1} converges asymptotically to the solution of the following Fokker-Planck equation in the forward-peaked limit \cite{pomraning1}:
\begin{equation}
\label{fp1}
\mu\frac{\partial \psi}{\partial x} + \sigma_a \psi = \frac{\sigma_{tr}}{2}\frac{\partial }{\partial \mu} (1-\mu^2) \frac{\partial \psi}{\partial \mu} + Q\,,
\end{equation}
where $\sigma_{tr}= \sigma_{s,0} -\sigma_{s,1}$ is the momentum transfer cross section and $\sigma_a = \sigma_t-\sigma_{s,0}$ is the macroscopic absorption cross section.

Source Iteration \cite{adamslarsen} is generally used to solve \cref{transport1}, which can be rewritten in operator notation:
\begin{equation}
\label{si1}
\mathcal{L} \psi^{m+1} = \mathcal{S} \psi^{m} + Q\,,
\end{equation}
where 
\begin{equation}
\mathcal{L} = \mu \frac{\partial}{\partial x} + \sigma_t,
   \quad
\mathcal{S} = \sum_{l=0}^L \frac{(2l+1)}{2} P_l(\mu) \sigma_{s,l} \int_{-1}^{1}d\mu P_l(\mu) ,
\label{trans1}
\end{equation}
and $m$ is the iteration index.
This equation is solved iteratively until a tolerance criterion is met. The FP approximation shown in \cref{fp1} can be used to accelerate the convergence of \cref{transport1}.

\subsection{FPSA: Fokker-Planck Synthetic Acceleration}\label{FPSA}

In the FPSA scheme \cite{JapanFPSA,japanDiss}, the FP approximation is used as a preconditioner to synthetically accelerate convergence when solving \cref{transport1} (cf. \cite{adamslarsen} for a detailed description of synthetic acceleration).
When solving \cref{si1}, the angular flux at each iteration $m$ has an error associated with it.
FPSA systematically follows a predict, correct, iterate scheme.
A transport sweep, one iteration in \cref{si1}, is made for a prediction.
The FP approximation is used to correct the error in the prediction, and this iteration is performed until a convergence criterion is met.
The equations used are:
\begin{subequations}
\label{fpsaeq}
\begin{align}
\label{predict}
\mathrm{Predict}&: \mathcal{L} \psi^{m+\frac{1}{2}} = \mathcal{S} \psi^{m} + Q\,,\\
\label{correct}
\mathrm{Correct}&: \psi^{m+1} =  \psi^{m+\frac{1}{2}} + \mathcal{P}^{-1} \mathcal{S} \left( \psi^{m+\frac{1}{2}} -  \psi^{m}\right),
\end{align}
\end{subequations}
where we define $\mathcal{P}$ as
\begin{equation}
\label{FPSAsi1}
\mathcal{P} = \mathcal{A}-\mathcal{F} =\underbrace{\left(\mu\frac{\partial}{\partial x} + \sigma_a\right)}_\mathcal{A} - \underbrace{\left(\frac{\sigma_{tr}}{2}\frac{\partial }{\partial \mu} (1-\mu^2) \frac{\partial }{\partial \mu}\right)}_\mathcal{F},
\end{equation}
In this synthetic acceleration method, the FP approximation is used to correct the error in each iteration of the high-order (HO) equation (\ref{predict}). 
Therefore, there is no consistency between the angular moments of the flux in the HO and low-order (LO) equations.

\subsection{NFPA: Nonlinear Fokker-Planck Acceleration}\label{NFPA}

Similar to FPSA, NFPA uses the FP approximation to accelerate the convergence of the solution.
We introduce the additive term $\hat{D}_F$ to \cref{fp1}, obtaining the modified FP equation
\begin{equation}
\label{mfp1}
\mu\frac{\partial \psi}{\partial x} + \sigma_a \psi = \frac{\sigma_{tr}}{2}\frac{\partial }{\partial \mu} (1-\mu^2) \frac{\partial \psi}{\partial \mu} + \hat{D}_F + Q\,.
\end{equation}
The role of $\hat{D}_F$ is to force the transport and modified FP equations to be consistent.
Subtracting \cref{mfp1} from \cref{transport1} and rearranging, we obtain the consistency term
\begin{equation}
\label{dfp}
\hat{D}_F = \sum_{l=0}^L \frac{(2l+1)}{2} P_l \sigma_l \phi_l - \frac{\sigma_{tr}}{2}\frac{\partial}{\partial \mu} (1-\mu^2) \frac{\partial \psi}{\partial \mu} - \sigma_{s,0} \psi\,.
\end{equation}

The NFPA method is given by the following equations:
\begin{subequations}\label[pluraleq]{holocons}
\begin{align}
\label{HO1}
\text{HO}&: \mu\frac{\partial \psi_{HO}}{\partial x} + \sigma_t \psi_{HO} = \sum_{l=0}^L \frac{(2l+1)}{2} P_l \sigma_l \phi_{l, LO} + Q\,,\\
\label{LO11}
\text{LO}&: \mu\frac{\partial \psi_{LO}}{\partial x} + \sigma_a \psi_{LO} = \frac{\sigma_{tr}}{2}\frac{\partial }{\partial \mu} (1-\mu^2) \frac{\partial \psi_{LO}}{\partial \mu} + \hat{D}_F + Q\,,\\
\label{con1}
\text{Consistency term}&: \hat{D}_F = \sum_{l=0}^L \frac{(2l+1)}{2} P_l \sigma_l \phi_{l, HO}^m - \frac{\sigma_{tr}}{2}\frac{\partial }{\partial \mu} (1-\mu^2) \frac{\partial \psi_{HO}}{\partial \mu} - \sigma_{s,0} \psi_{HO}\,,
\end{align}
\end{subequations}
where $\psi_{HO}$ is the angular flux obtained from the HO equation and $\psi_{LO}$ is the angular flux obtained from the LO equation.
The nonlinear HOLO-plus-consistency system given by \cref{holocons} can be solved using any nonlinear solution technique \cite{kelley}. Note that the NFPA scheme returns a FP equation that is consistent with HO transport. 
Moreover, this modified FP equation accounts for large-angle scattering which the standard FP equation does not. 
The LO equation (\ref{fp1}) can then be integrated into multiphysics models in a similar fashion to standard HOLO schemes \cite{patelFBR}. To solve the HOLO-plus-consistency system above, we use Picard iteration \cite{kelley}:
\begin{subequations}
\begin{align}
\label{H1}
\text{Transport Sweep for HO}&:
\mathcal{L} \psi_{HO}^{k+1} = \mathcal{S} \psi_{LO}^{k} + Q, \\
\label{L1}
\text{Evaluate Consistency Term}&: \hat{D}_F^{k+1} = \left(\mathcal{S} - \mathcal{F} - \sigma_{s,0}\mathcal{I}\right) \psi_{HO}^{k+1}, \\
\label{c1}
\text{Solve LO Equation}&: \psi_{LO}^{k+1} = \mathcal{P}^{-1} \left(\hat{D}_F^{k+1} + Q\right), 
\end{align}
\end{subequations}
where $\mathcal{L}$ and $\mathcal{S}$ are given in \cref{trans1}, $\mathcal{P}$ and $\mathcal{F}$ are given in \cref{FPSAsi1}, $\mathcal{I}$ is the identity operator, and $k$ is the iteration index.
Iteration is done until a convergence criterion is met.

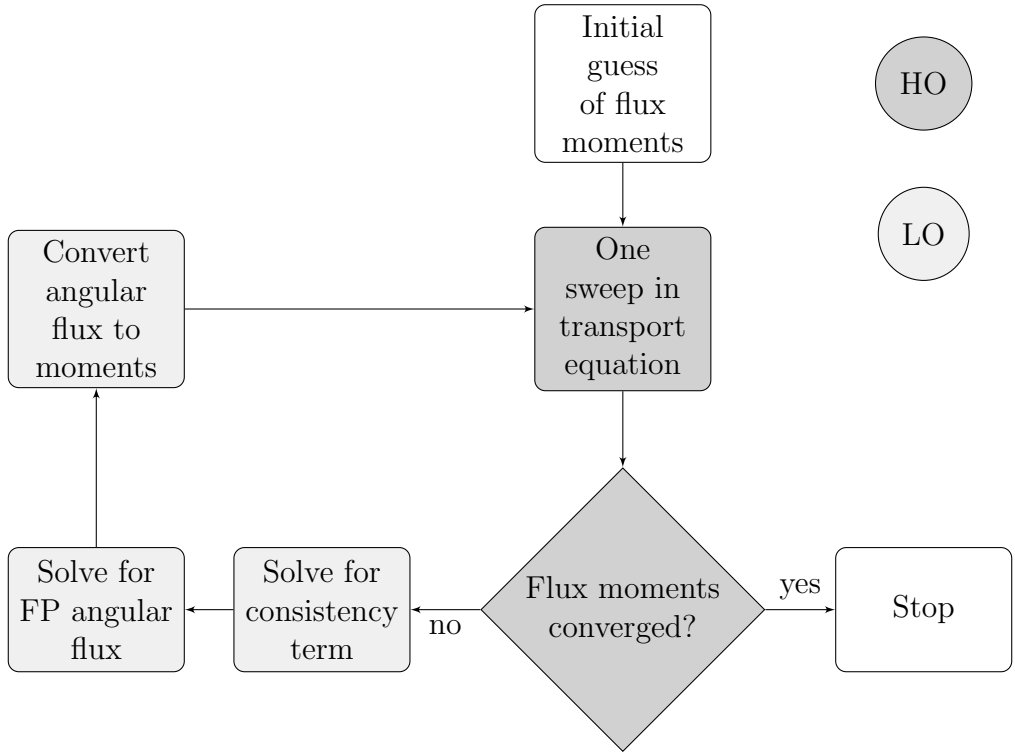
\begin{figure}[H]
\centering
\begin{tikzpicture}[node distance = 3cm, auto]
    % Place nodes
    \node [block] (init) {Initial guess of flux moments};
    \node [cloud_HO, right of=init, node distance=4cm] (HOm) {HO};
    \node [cloud_LO, below of=HOm, node distance=2cm] (LOm) {LO};
    \node [HO, below of=init] (transport) {One sweep in transport equation};
    \node [decision, below of=transport,node distance=4cm] (decide) {Flux moments converged?};
    \node [LO, left of=decide, node distance=4cm] (dterm) {Solve for consistency term};
    \node [LO, left of=dterm, node distance=3cm] (MFP) {Solve for FP angular flux};
    \node [LO, above of=MFP, node distance=4cm] (moments) {Convert angular flux to moments};
    \node [block, right of=decide, node distance=4cm] (stop) {Stop};
    % Draw edges
    \path [line] (init) -- (transport);
    \path [line] (transport) -- (decide);
    \path [line] (decide) -- node {no} (dterm);
    \path [line] (dterm) -- (MFP);
    \path [line] (MFP) -- (moments);
    \path [line] (moments) -- (transport);
    \path [line] (decide) -- node {yes}(stop);
\end{tikzpicture}
\caption{NFPA algorithm}
\label{Nalgorithm}
\end{figure}

The main advantage of setting up the LO equation in this fashion is that the stiffness matrix for LO needs to be setup and inverted \textit{only once}, just as with FPSA \cite{JapanFPSA, japanDiss}. This has a large impact on the method's performance.
A flowchart of this algorithm is shown in \cref{Nalgorithm}.

\section{Numerical Experiments}\label{sec3}

In \cref{sec31} we describe the discretization methods used to implement the algorithms.
In \cref{sec32} we provide numerical results for 2 different choices of source $Q$ and boundary conditions.
For each choice we solve the problem using 3 different scattering kernels, applying 3 different choices of parameters for each kernel.
We provide NFPA numerical results for these 18 cases and compare them against those obtained from FPSA and other standard methods.

All numerical experiments were performed using MATLAB.
Runtime was tracked using the tic-toc functionality \cite{matlab17}, with
only the solver runtime being taken into consideration in the comparisons.
A 2017 MacBook Pro with a 2.8 GHz Quad-Core Intel Core i7 and 16 GB of RAM was used for all simulations.

\subsection{Discretization}\label{sec31}

The Transport and FP equations were discretized using linear discontinuous finite element discretization in space \cite{mpd1}, and discrete ordinates (S$_N$) in angle \cite{landm}.
The Fokker-Planck operator $\mathcal{F}$ was discretized using moment preserving discretization (MPD) \cite{mpd1}.
Details of the derivation of the linear discontinuous finite element discretization can be seen in \cite{japanDiss,martin}.
The finite element discretization for the  Fokker-Planck equation follows the same derivation.

A brief review for the angular discretization used for the FP equation is given below.
First, we use Gauss-Legendre quadrature to discretize the FP equation in angle:
\begin{equation}
\mu_n\frac{\partial \psi_n(x)}{\partial x} + \sigma_a \psi_n(x) - \frac{\sigma_{tr}}{2}\nabla^2_n \psi_n(x) =  Q_n(x),
\end{equation}
for $n=1,..,N$.
Here, $\nabla^2_n$ term is the discrete form of the angular Laplacian operator evaluated at angle $n$.

The MPD scheme is then shown as
\begin{equation}
\nabla^2_n \psi_n = M \psi_n = V^{-1} L V \psi_n,
\end{equation}
where $M$ is the MPD discretized operator defined by
\begin{subequations}
\begin{equation}
V_{i,j} = P_{i-1}(\mu_j)w_j,
\end{equation}
and 
\begin{equation}
L_{i,j} = -i(i-1),
\end{equation}
\end{subequations}
for $i,j=1,...,N$.
Here, $P_l(\mu_j)$ are the Legendre polynomials evaluated at each angle $\mu_j$ and $w_j$ are the respective weights.
$M$ is defined as a (N x N) operator for a vector of $N$ angular fluxes $ \psi(x)$, at spatial location $x$. 

In summary, if we write the FP equation as
\begin{equation}
\mathcal{H} \frac{\partial \psi}{\partial x}(x) + \sigma_a \psi(x) - \mathcal{F} \psi(x) =  Q(x),
\end{equation}
then $\mathcal{H}$ is Diag$(\mu_n)$ for $n=1,...,N$, $Q(x)$ is a vector of source terms $Q_n(x)$, and $\mathcal{F}$ is represented by $\frac{\sigma_{tr}}{2}M$.

\subsection{Numerical Results}\label{sec32}

It is shown that for slowly converging problems, typical convergence methods like $L_\infty$ suffer from false convergence \cite{adamslarsen}.
To work around this issue, the criterion is modified to use information about the current and previous iteration:
\begin{equation}
\label{falseconverge}
\frac{|| \phi^{m}_0(x) - \phi^{m-1}_0(x) ||_2}{1-\frac{|| \phi^{m+1}_0(x) - \phi^{m}_0(x) ||_2}{|| \phi^{m}_0(x) - \phi^{m-1}_0(x) ||_2}} < 10^{-8}.
\end{equation}

Two problems were tested using 200 spatial cells, $X$ = 400, $\sigma_a = 0$, $L$ = 15, and $N$ = 16.
Problem 1 has vacuum boundaries and a homogeneous isotropic source $Q$ for $0 < x < X$.
Problem 2 has no internal source and an incoming beam at the left boundary. The source and boundary conditions used are shown in \cref{parameters}. 
\begin{table}[H]
\begin{center}
\scalebox{0.9}{
\begin{tabular}{c | c | c} \hline 
& Problem 1 & Problem 2 \\ \hline \hline
Q(x) & 0.5 & 0 \\
$\psi_L$ & 0 & $\delta(\mu - \mu_N)$  \\
$\psi_R$ & 0 & 0 \\
\end{tabular}}
\end{center}
\caption{Problem Parameters}
\label{parameters} 
\end{table} 
We consider three scattering kernels in this paper: Screened Rutherford \cite{pomraning1}, Exponential \cite{pomraning2}, and Henyey-Greenstein \cite{HGK}.
Three cases for each kernel were tested.
The results obtained with NFPA are compared with those obtained using GMRES, DSA, and FPSA with the MPD scheme.

\subsubsection{SRK: Screened Rutherford Kernel}

\begin{figure}[t]
\begin{center}
  \includegraphics[scale=0.1,angle=0]{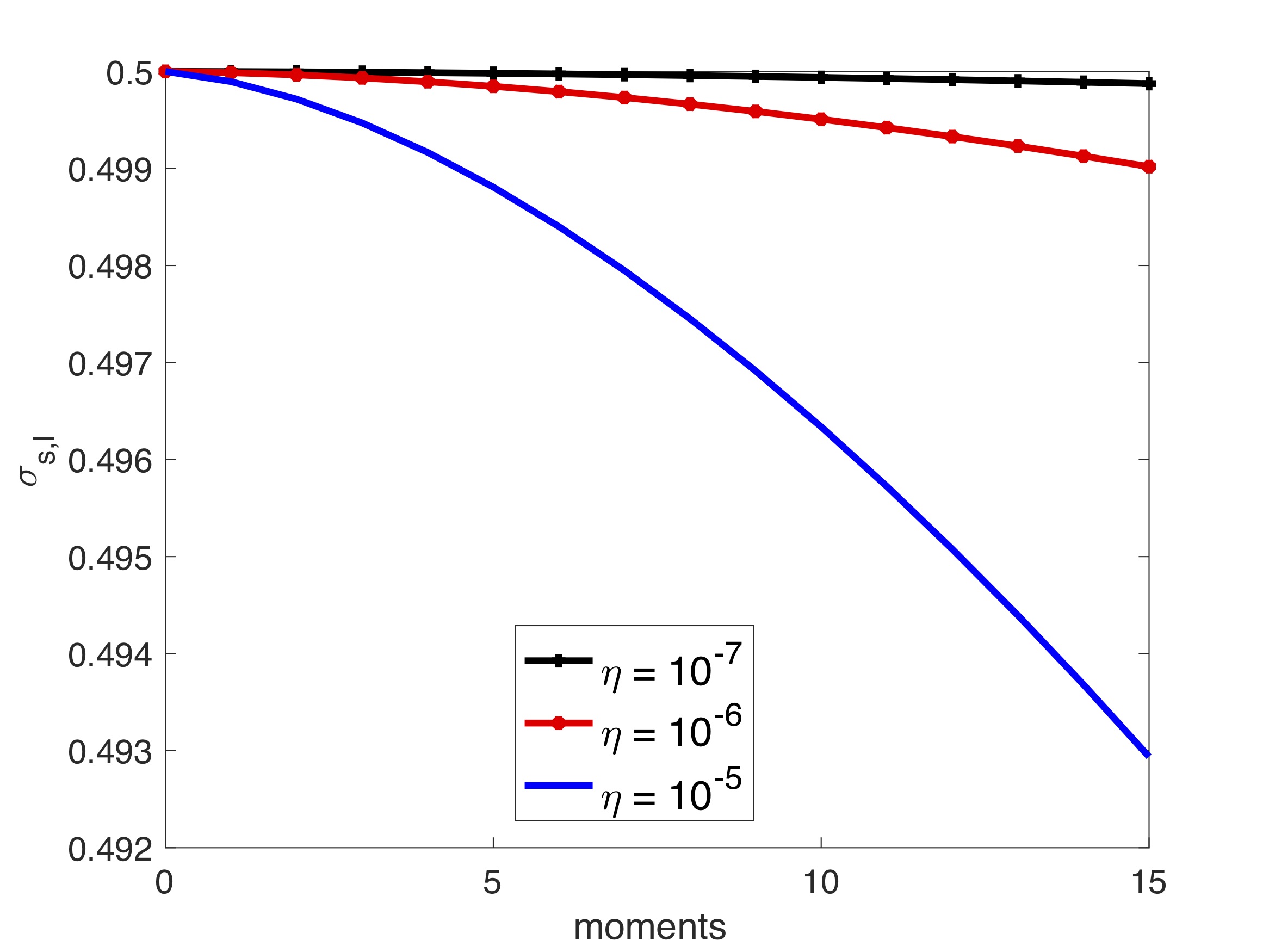}
  \caption{Screened Rutherford Kernels}
  \label{SRK}
\end{center}
\end{figure}

\begin{figure}[H]
    \centering
    \subfloat[Problem 1]{{\includegraphics[width=7cm]{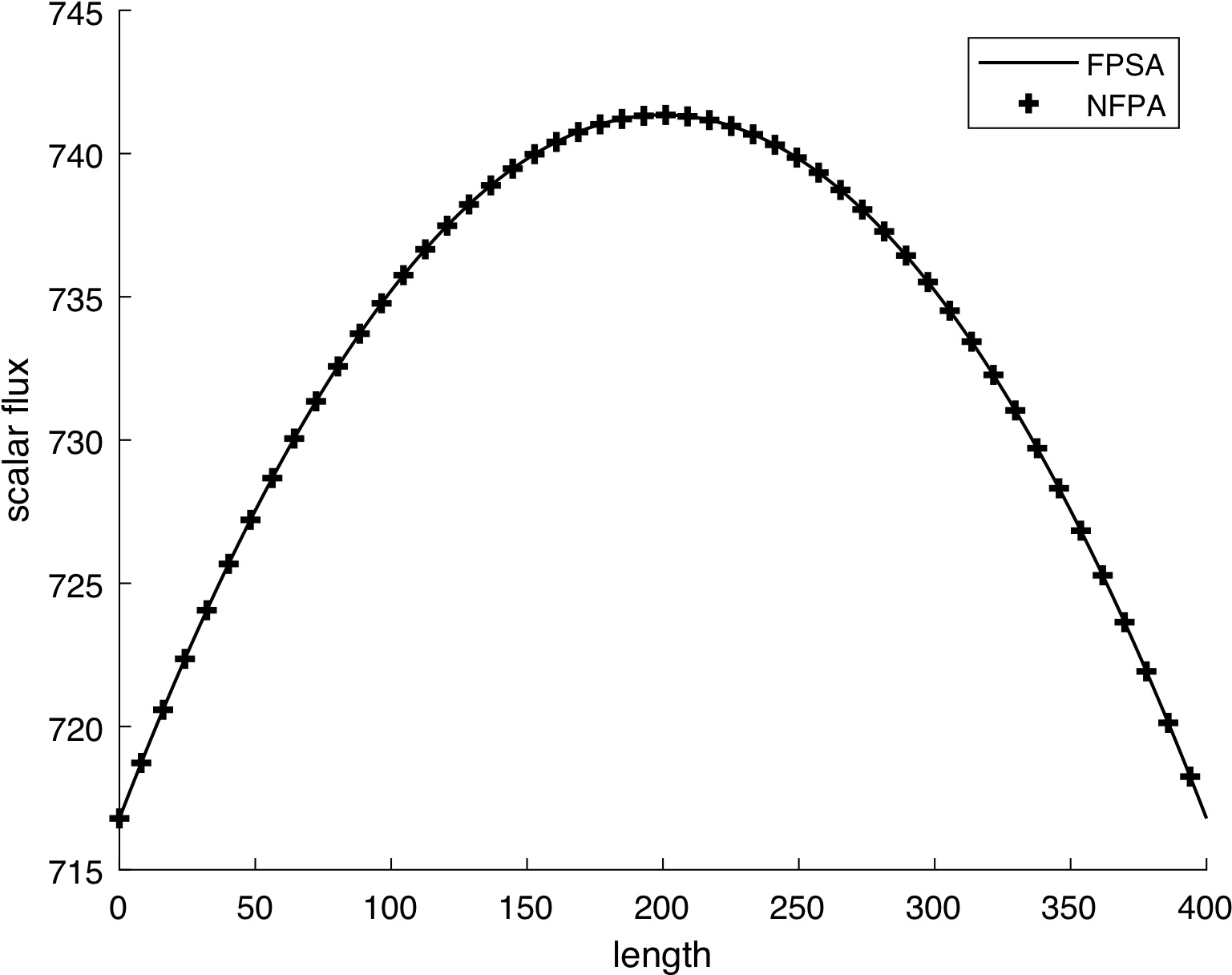} }}
    \qquad
    \subfloat[Problem 2]{{\includegraphics[width=7cm]{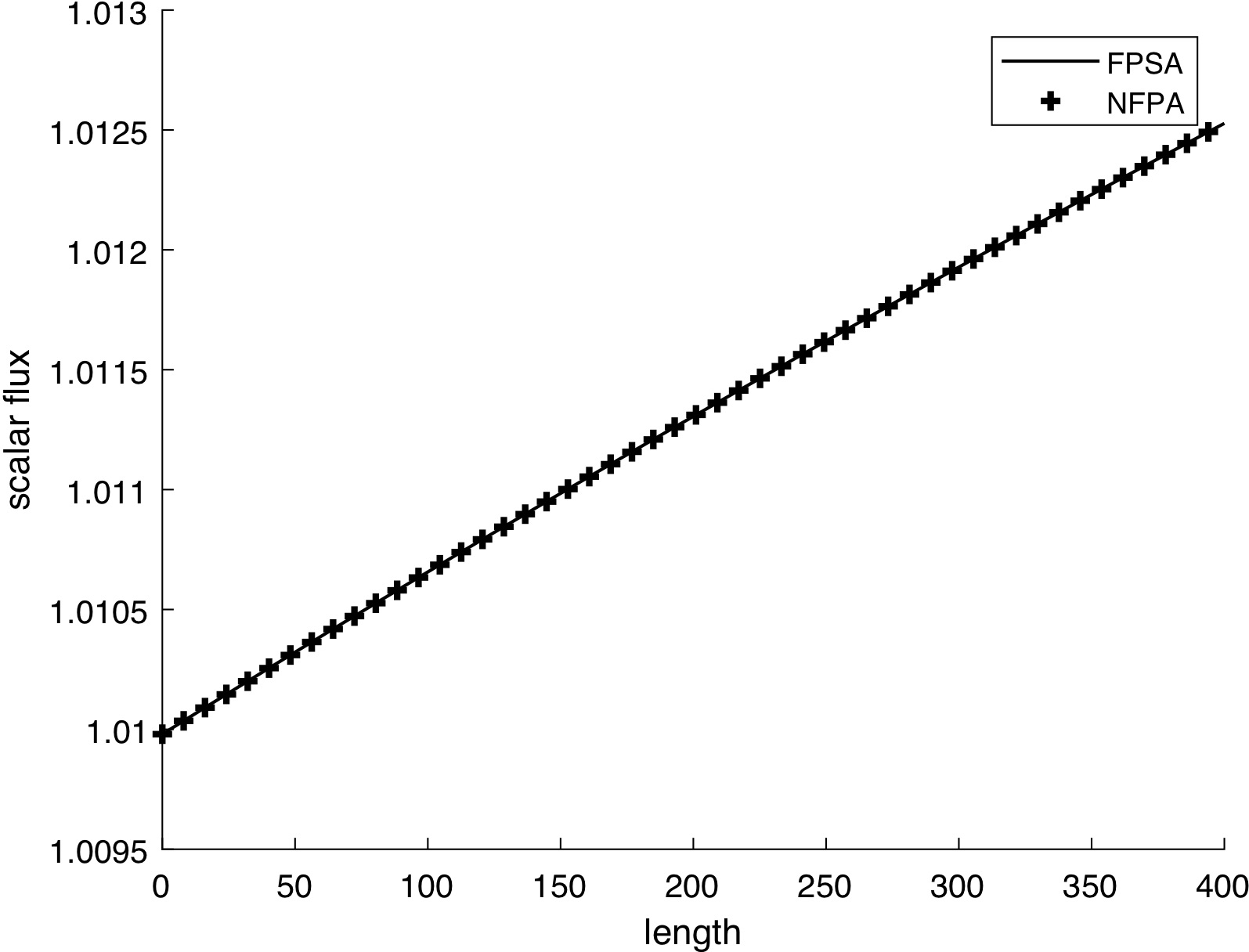} }}
    \caption{Results for SRK Problems with $\eta = 10^{-7}$}
    \label{SRK_plots}
\end{figure}

The Screened Rutherford Kernel \cite{pomraning1, JapanFPSA} is a widely used scattering kernel for modeling scattering behavior of electrons \cite{SRK}.
The kernel depends on the parameter $\eta$, such that
\begin{equation}
\sigma^{SRK}_{s,l} = \sigma_s  \int_{-1}^{1} d\mu P_l(\mu) \frac{\eta (\eta+1)}{(1+2\eta-\mu)^2}.
\end{equation}
The SRK has a valid FP limit as $\eta$ approaches 0 \cite{patelFBR}. Three different values of $\eta$ were used to generate the scattering kernels shown in \cref{SRK}.
GMRES, DSA, FPSA, and NFPA all converged to the same solution for problems 1 and 2. \Cref{SRK_plots} shows the solutions for SRK with $\eta = 10^{-7}$.

\begin{table}[H]
\begin{center}
\scalebox{0.8}{
\begin{tabular}{c || c || c || c} \hline 
Parameter & Solver & Runtime (s) & Iterations \\ \hline \hline
\multirow{4}{*}{$\eta = 10^{-5}$} & GMRES & 98.8 & 12 \\
& DSA & 2380 & 53585 \\
& FPSA & 1.21 & 26 \\
%& NFPA-WFD & 2.21 & 11 \\
& NFPA & 1.39 & 26 \\ \hline 
\multirow{4}{*}{$\eta = 10^{-6}$} & GMRES & 208 & 84 \\
& DSA & 3040 & 69156 \\
& FPSA & 0.747 & 16 \\
%& NFPA-WFD & 1.43 & 8 \\
& NFPA & 0.857 & 16 \\ \hline 
\multirow{4}{*}{$\eta = 10^{-7}$} & GMRES & 174 & 124 \\
& DSA & 3270 & 73940 \\
& FPSA & 0.475 & 10 \\
%& NFPA-WFD & 1.17 & 6 \\
& NFPA & 0.542 & 10 \\ \hline
\end{tabular}}
\end{center}
\caption{Runtime and Iteration Counts for Problem 1 with SRK}
\label{SRKresults1} 
\end{table}
\begin{table}[H]
\begin{center}
\scalebox{0.8}{
\begin{tabular}{c || c || c || c} \hline 
Parameter & Solver & Runtime (s) & Iterations \\ \hline \hline
\multirow{4}{*}{$\eta = 10^{-5}$} & GMRES & 52.4 & 187 \\
& DSA & 1107 & 25072 \\
& FPSA & 0.953 & 20 \\
%& NFPA-WFD & 2.56 & 13 \\
& NFPA & 1.14 & 20 \\ \hline 
\multirow{4}{*}{$\eta = 10^{-6}$} & GMRES & 108 & 71 \\
& DSA & 1434 & 32562  \\
& FPSA & 0.730 & 14 \\
%& NFPA-WFD & 2.42 & 12 \\
& NFPA & 0.857 & 14 \\ \hline 
\multirow{4}{*}{$\eta = 10^{-7}$} & GMRES & 94.1 & 185 \\
& DSA & 1470 & 33246 \\
& FPSA & 0.438 & 8 \\
%& NFPA-WFD & 1.61 & 8 \\
& NFPA & 0.484 & 8 \\ \hline  
\end{tabular}}
\end{center}
\caption{Runtime and Iteration Counts for Problem 2 with SRK}
\label{SRKresults2} 
\end{table}

The results of all solvers are shown in \cref{SRKresults1,SRKresults2}.
We see that NFPA and FPSA tremendously outperform GMRES and DSA in runtime for all cases.
FPSA is a simpler method than NFPA, requiring less calculations per iteration; therefore, it is expected that it outperforms NFPA in runtime.
We see a reduction in runtime and iterations for FPSA and NFPA as the FP limit is approached, with DSA and GMRES requiring many more iterations by comparison as $\eta$ approaches 0.

\begin{figure}[H]
\begin{center}
  \includegraphics[scale=0.15,angle=0]{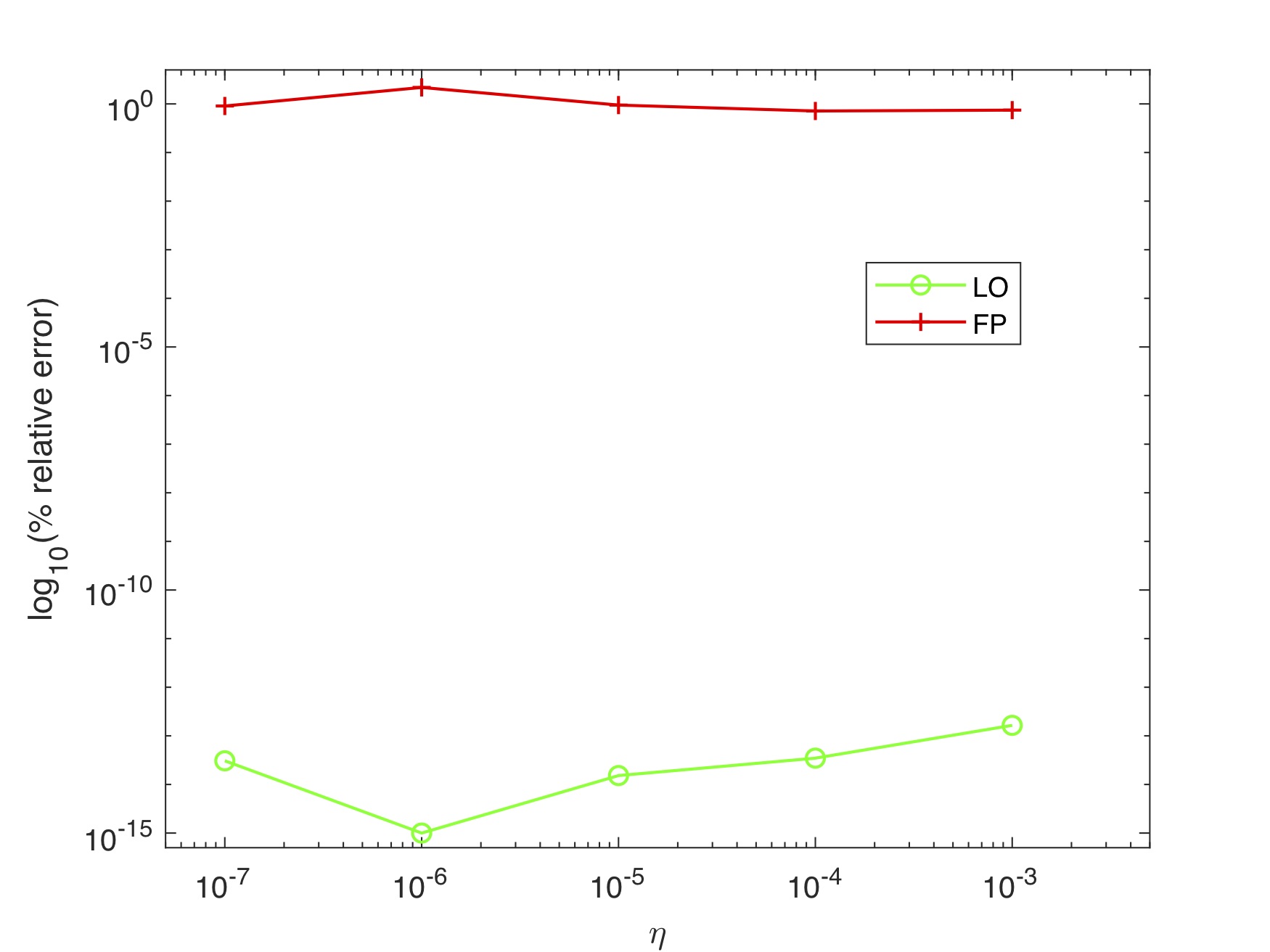}
  \caption{Log Scale of $\%$ Relative Error vs $\eta$ for Problem 1 at the Center of the Slab with SRK}
  \label{momcomp}
\end{center}
\end{figure}

An advantage that NFPA offers is that the angular moments of the flux in the LO equation will remain consistent with those of the transport equation even as a problem becomes less forward-peaked.
On the other hand, the moments found using only the FP equation and source iteration lose accuracy.
To illustrate this, Problem 1 was tested using different Screened Rutherford Kernels with increasing $\eta$ parameters.
The percent errors (relative to the transport solution) for the scalar flux obtained with the LO equation and with the standard FP equation at the center of the slab are shown in \cref{momcomp}.
It can be seen that the percent relative errors in the scalar flux of the FP solution is orders of magnitude larger than the error produced using the LO equation.
The same trend can be seen when using the exponential and Henyey-Greenstein kernels. 

\subsubsection{EK: Exponential Kernel}

The exponential kernel \cite{pomraning2, JapanFPSA} is a fictitious kernel made for problems that have a valid Fokker-Planck limit \cite{pomraning1}.
The zero$^{\text{th}}$ moment, $\sigma^{EK}_{s,0}$, is chosen arbitrarily; we define $\sigma^{EK}_{s,0}$ as the same zero$^{\text{th}}$ moment from the SRK.
The $\Delta$ parameter determines the kernel: the first and second moments are given by 
\begin{subequations}
\begin{align}
\sigma^{EK}_{s,1} &= \sigma^{EK}_{s,0} (1-\Delta),\\
\sigma^{EK}_{s,2} &= \sigma^{EK}_{s,0} (1-3\Delta+3\Delta^2),
\end{align}
and the relationship for $l\geq 3$ is
\begin{equation}
\sigma^{EK}_{s,l} = \sigma^{EK}_{s,l-2} - \Delta(2l+1) \sigma^{EK}_{s,l-1}.
\end{equation}
\end{subequations}
As $\Delta$ is reduced, the scattering kernel becomes more forward-peaked.

\begin{figure}[t]
\begin{center}
  \includegraphics[scale=0.1,angle=0]{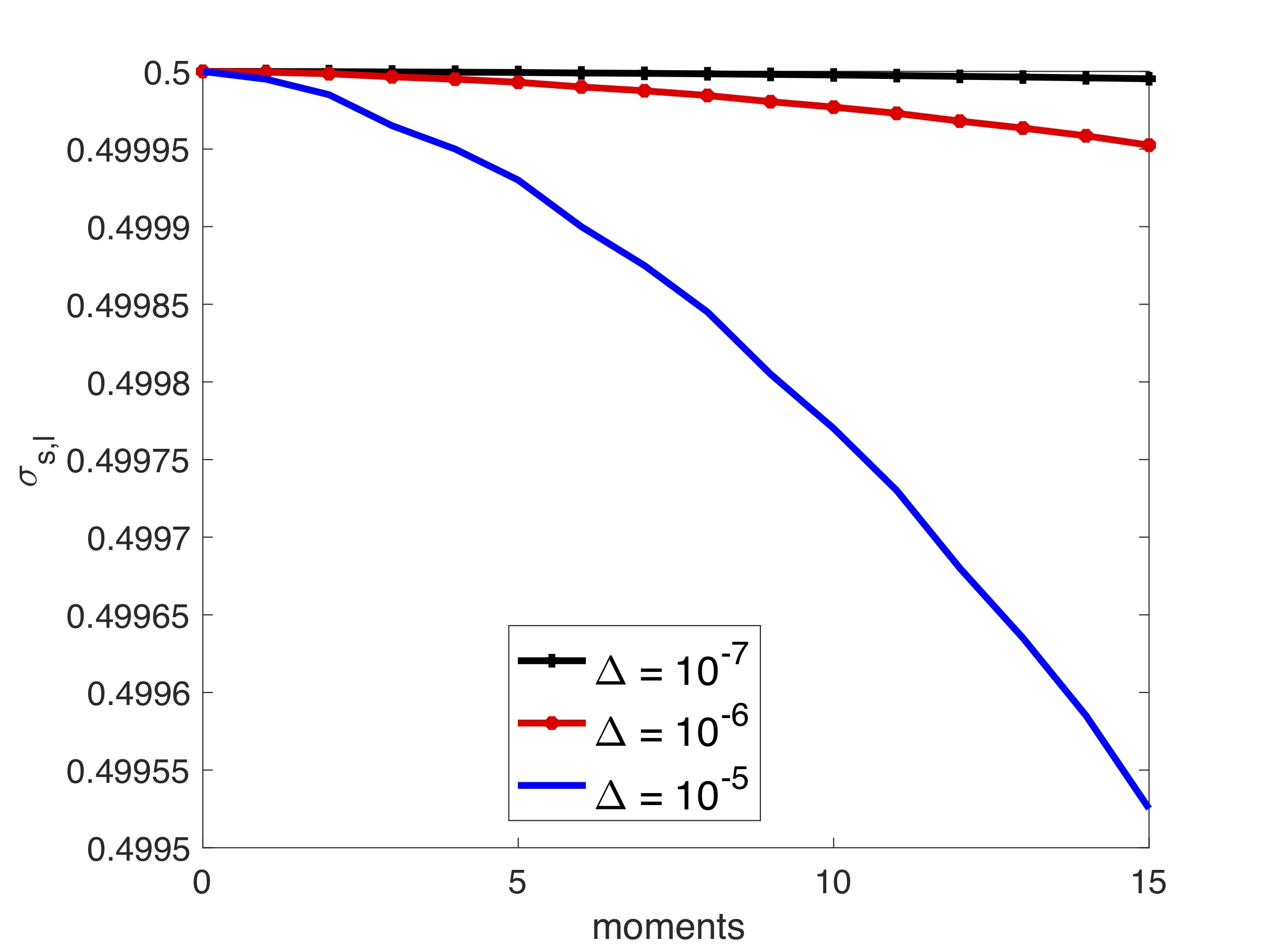}
  \caption{Exponential Kernels}
  \label{EXP}
\end{center}
\end{figure}

\begin{figure}[H]
    \centering
    \subfloat[Problem 1]{{\includegraphics[width=7cm]{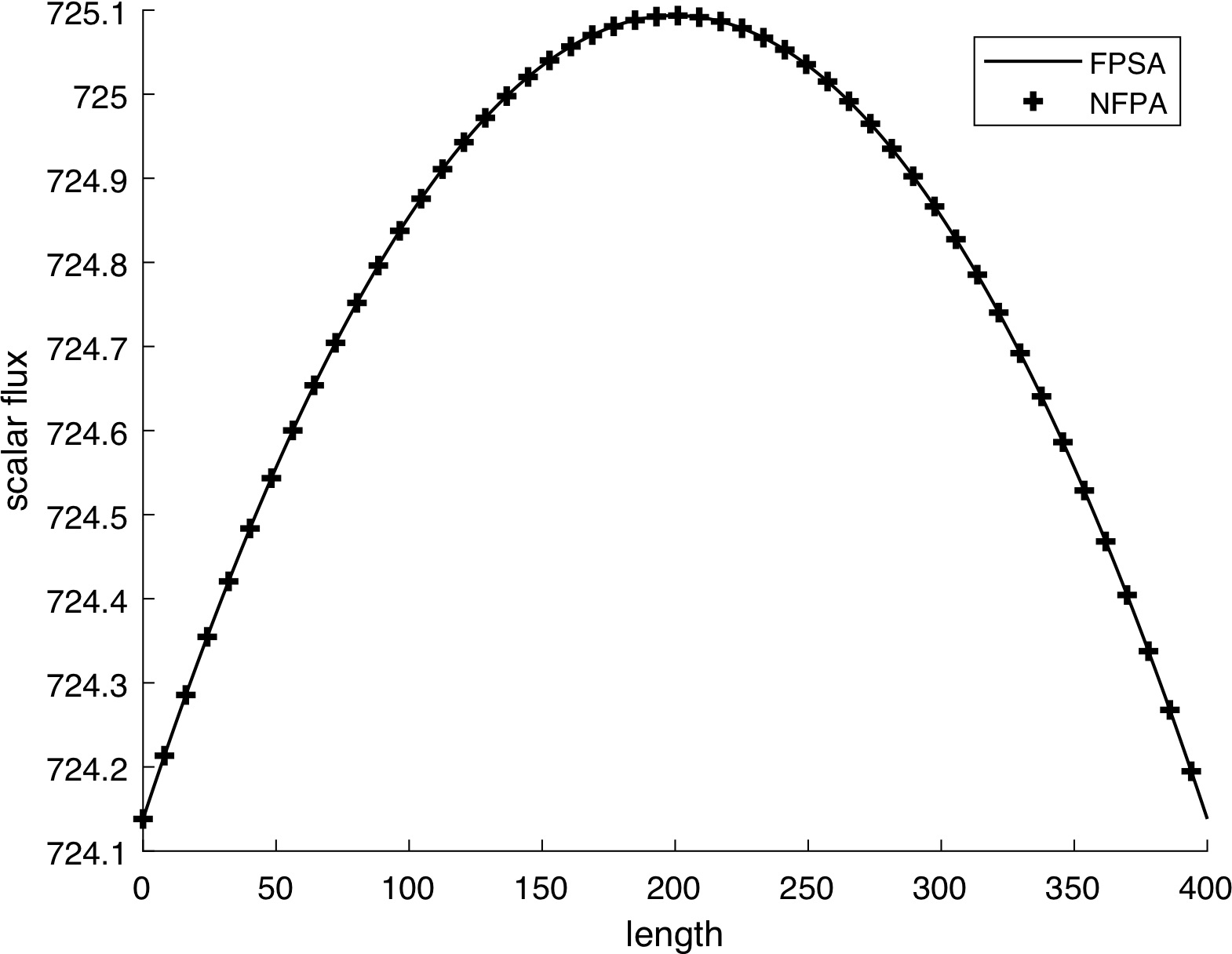} }}
    \qquad
    \subfloat[Problem 2]{{\includegraphics[width=7cm]{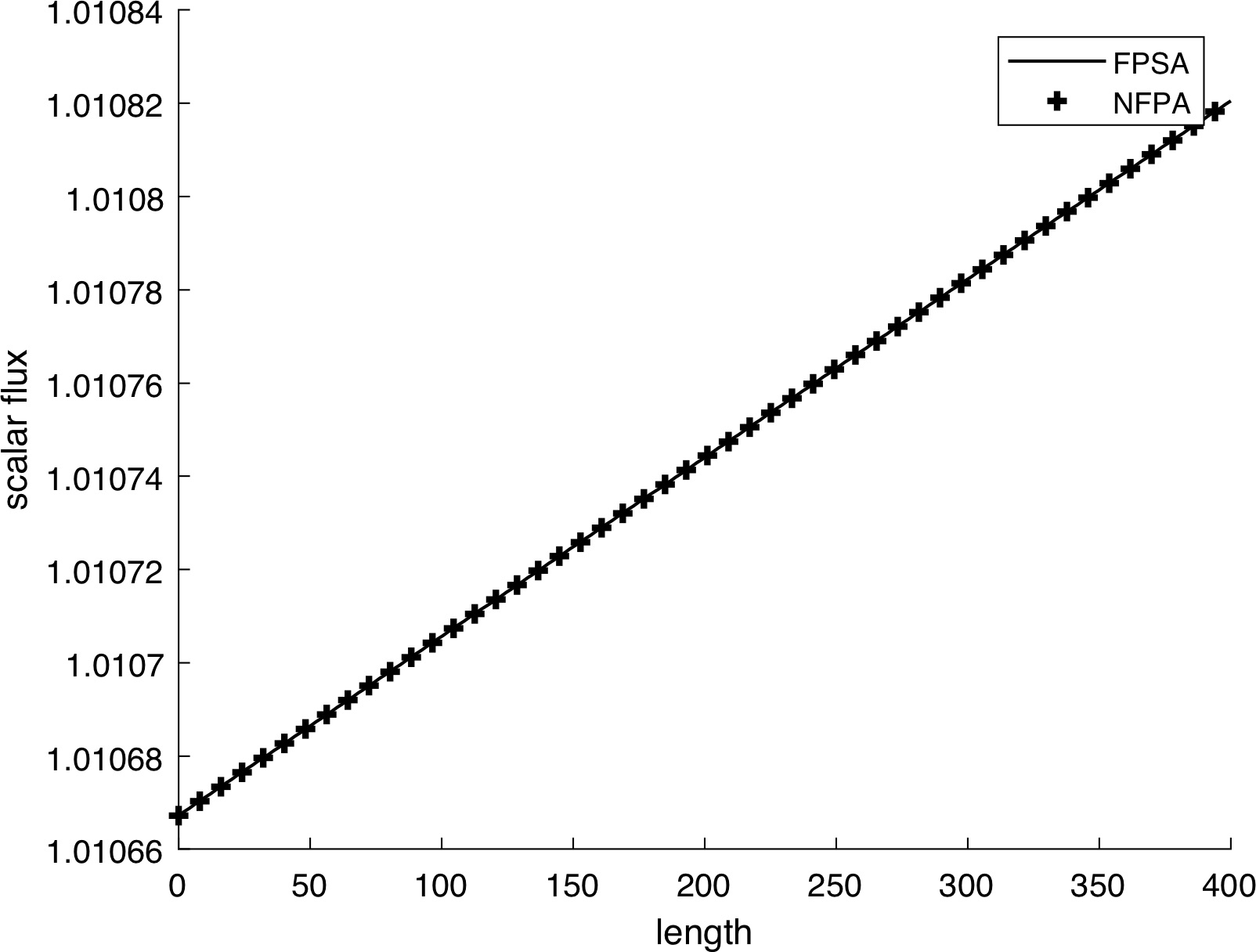} }}
    \caption{Results for EK Problems with $\Delta = 10^{-7}$}
    \label{EK_plots}
\end{figure}

The EK has a valid FP limit as $\Delta$ approaches 0 \cite{patelFBR}.
Three different values of $\Delta$ were used to generate the scattering kernels shown in \cref{EXP}.
The generated scattering kernels are shown in \cref{EXP}.
GMRES, DSA, FPSA, and NFPA all converged to the same solution for problems 1 and 2.
\Cref{EK_plots} shows the solutions for EK with $\Delta = 10^{-7}$.

The runtimes and iterations for GMRES, DSA, FPSA, and NFPA are shown in \cref{Expresults1,Expresults2}.
We see a similar trend with the EK as seen with SRK.
Smaller $\Delta$ values lead to a reduction in runtime and iterations for NFPA and FPSA, which greatly outperform DSA and GMRES in both categories.

\begin{table}[h]
\begin{center}
\scalebox{0.8}{
\begin{tabular}{c || c || c || c} \hline 
Parameter & Solver & Runtime (s) & Iterations \\ \hline \hline
\multirow{4}{*}{$\Delta = 10^{-5}$} & GMRES & 196 & 142 \\
& DSA & 3110 & 70140 \\
& FPSA & 0.514 & 11 \\ 
%& NFPA-WFD & 1.56 & 8 \\
& NFPA & 0.630 & 11 \\\hline 
\multirow{4}{*}{$\Delta = 10^{-6}$} & GMRES & 156 & 132 \\
& DSA & 3120 & 70758 \\
& FPSA & 0.388 & 7 \\ 
%& NFPA-WFD & 1.16 & 6 \\
& NFPA & 0.393 & 7 \\ \hline 
\multirow{4}{*}{$\Delta = 10^{-7}$} & GMRES & 81 & 127 \\
& DSA & 3120 & 70851  \\
& FPSA & 0.292 & 6 \\ 
%& NFPA-WFD & 0.977 & 5 \\
& NFPA & 0.318 & 6 \\ \hline
\end{tabular}}
\end{center}
\caption{Runtime and Iteration Counts for Problem 1 with EK}
\label{Expresults1} 
\end{table}
\begin{table}[h]
\begin{center}
\scalebox{0.8}{
\begin{tabular}{c || c || c || c} \hline 
Parameter & Solver & Runtime (s) & Iterations \\ \hline \hline
\multirow{4}{*}{$\Delta = 10^{-5}$} & GMRES & 110 & 73 \\
& DSA & 1455 & 33033 \\
& FPSA & 0.492 & 10 \\ 
%& NFPA-WFD & 2.41 & 12 \\
& NFPA & 0.613 & 10 \\ \hline 
\multirow{4}{*}{$\Delta = 10^{-6}$} & GMRES & 82.7 & 79 \\
& DSA & 1470 & 33309 \\
& FPSA & 0.358 & 7 \\ 
%& NFPA-WFD & 1.39 & 7 \\
& NFPA & 0.431 & 7 \\ \hline 
\multirow{4}{*}{$\Delta = 10^{-7}$} & GMRES & 56.8 & 90 \\
& DSA & 1470 & 33339 \\
& FPSA & 0.273 & 5 \\ 
%& NFPA-WFD & 0.992 & 5 \\
& NFPA & 0.319 & 5 \\ \hline  
\end{tabular}}
\end{center}
\caption{Runtime and Iteration Counts for Problem 2 with EK}
\label{Expresults2} 
\end{table}

\subsubsection{HGK: Henyey-Greenstein Kernel}

The Henyey-Greenstein Kernel \cite{HGK,JapanFPSA} is most commonly used in light transport in clouds.
It relies on the anisotropy factor $g$, such that
\begin{equation}
\sigma^{HGK}_{s,l} = \sigma_s g^l.
\end{equation}
As $g$ goes from zero to unity, the scattering shifts from isotropic to highly anisotropic.
\begin{figure}[H]
\begin{center}
  \includegraphics[scale=0.1,angle=0]{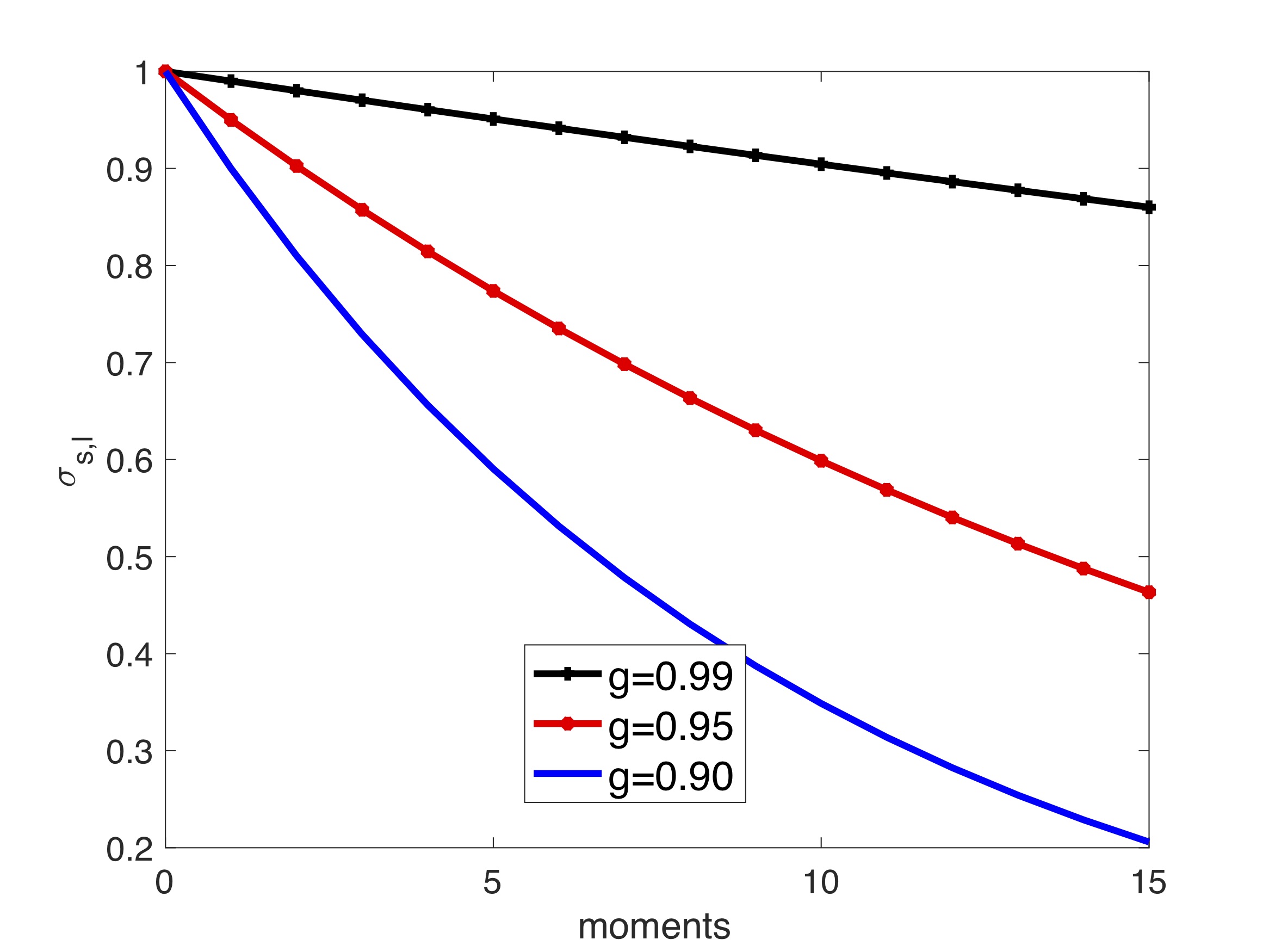}
  \caption{Henyey-Greenstein Kernels}
  \label{HGK}
\end{center}
\end{figure}
\begin{figure}[H]
    \centering
    \subfloat[Problem 1]{{\includegraphics[width=7cm]{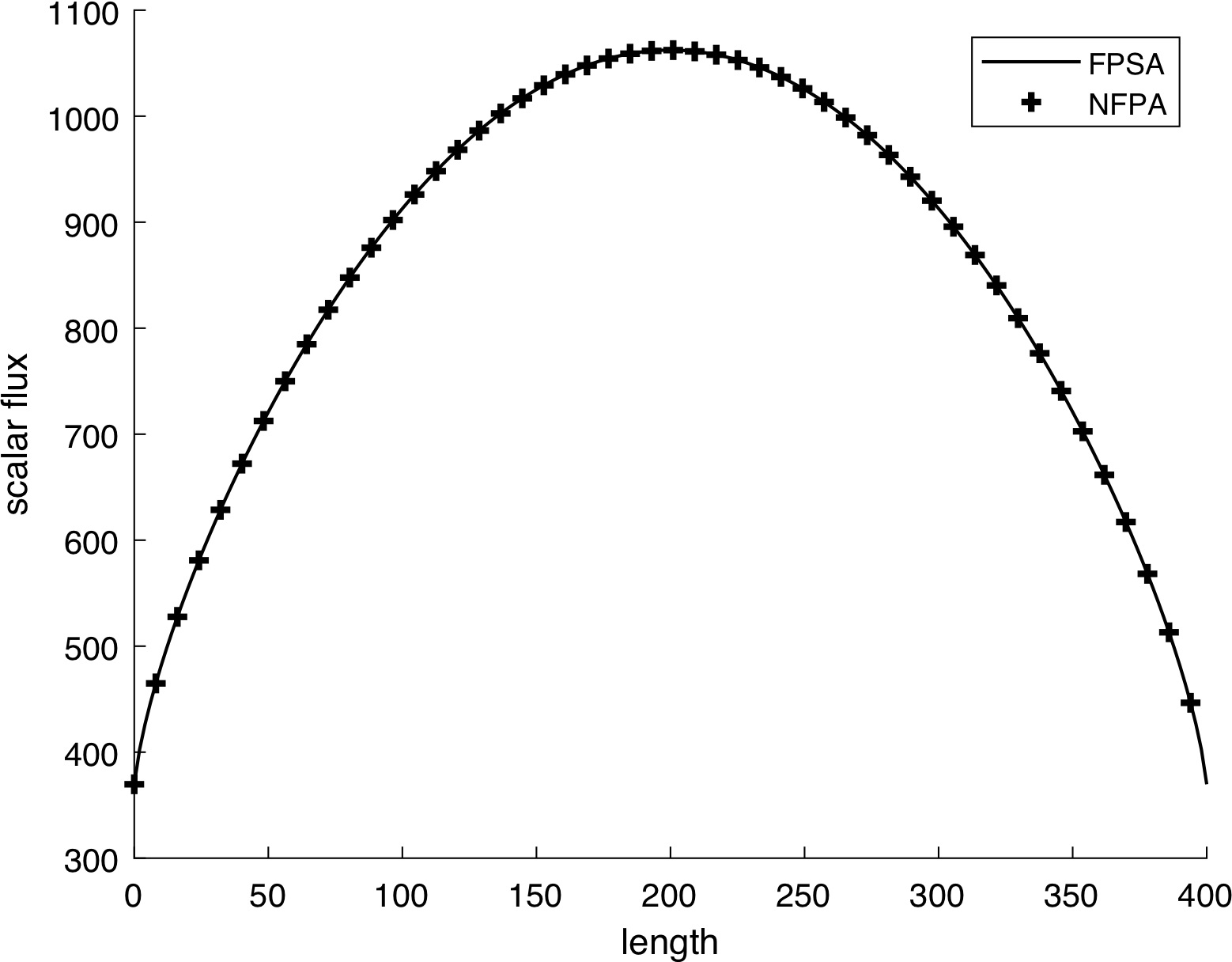} }}
    \qquad
    \subfloat[Problem 2]{{\includegraphics[width=7cm]{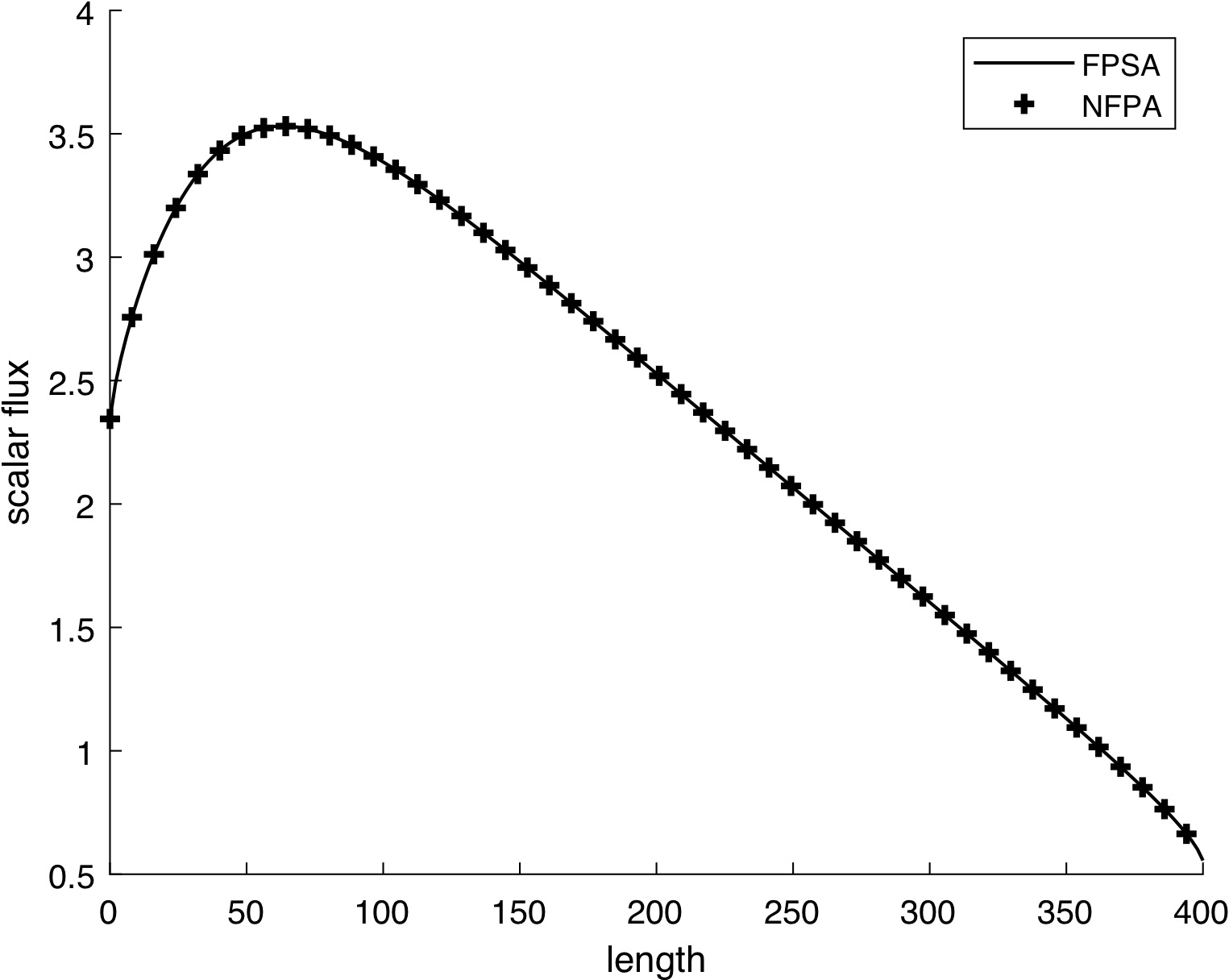} }}
    \caption{Results for HGK Problems with $g = 0.99$}
    \label{HGK_plots}
\end{figure}

The HGK does not have a valid FP limit \cite{patelFBR}.
The three kernels tested are shown in \cref{HGK}.
GMRES, DSA, FPSA, and NFPA all converged to the same solution for problems 1 and 2.
\Cref{HGK_plots} shows the solutions for HGK with $g = 0.99$.
The results of each solver are shown in \cref{HGKresults1,HGKresults2}. 
\begin{table}[h]
\begin{center}
\scalebox{0.8}{
\begin{tabular}{c || c || c || c} \hline 
Parameter & Solver & Runtime (s) & Iterations \\ \hline \hline
\multirow{4}{*}{$g=0.9$} & GMRES & 9.88 & 76 \\
& DSA & 24.5 & 554 \\
& FPSA & 1.50 & 32 \\ 
%& NFPA-WFD & 3.20 & 16 \\
& NFPA & 1.39 & 27 \\ \hline 
\multirow{4}{*}{$g=0.95$} & GMRES & 12.2 & 131 \\
& DSA & 47.7 & 1083 \\
& FPSA & 1.75 & 38 \\ 
%& NFPA-WFD & 4.03 & 20 \\
& NFPA & 1.83 & 35 \\ \hline 
\multirow{4}{*}{$g=0.99$} & GMRES & 40.0 & 27 \\
& DSA & 243 & 5530  \\
& FPSA & 3.38 & 74 \\ 
%& NFPA-WFD & 6.41 & 32 \\
& NFPA & 3.93 & 73 \\ \hline
\end{tabular}}
\end{center}
\caption{Runtime and Iteration Counts for Problem 1 with HGK}
\label{HGKresults1} 
\end{table}
\begin{table}[h]
\begin{center}
\scalebox{0.8}{
\begin{tabular}{c || c || c || c} \hline 
Parameter & Solver & Runtime (s) & Iterations \\ \hline \hline
\multirow{4}{*}{$g=0.9$} & GMRES & 24.3 & 135 \\
& DSA & 14.8 & 336  \\
& FPSA & 1.15 & 23 \\ 
%& NFPA-WFD & 4.66 & 23 \\
& NFPA & 1.35 & 24 \\ \hline 
\multirow{4}{*}{$g=0.95$} & GMRES & 31.3 & 107 \\
& DSA & 29.7 & 675 \\
& FPSA & 1.56 & 32 \\ 
%& NFPA-WFD & 6.29 & 31 \\
& NFPA & 1.90 & 33 \\ \hline 
\multirow{4}{*}{$g=0.99$} & GMRES & 41.4 & 126 \\
& DSA & 146 & 3345 \\
& FPSA & 3.31 & 67 \\ 
%& NFPA-WFD & 10.4 & 51 \\
& NFPA & 3.99 & 67 \\ \hline  
\end{tabular}}
\end{center}
\caption{Runtime and Iteration Counts for Problem 2 with HGK}
\label{HGKresults2} 
\end{table}

Here we see that NFPA and FPSA do not perform as well compared to their results for the SRK and EK.
Contrary to what happened in those cases, both solvers require more time and iterations as the problem becomes more anisotropic.
This is somewhat expected, due to HGK not having a valid Fokker-Planck limit.
However, both NFPA and FPSA continue to greatly outperform GMRES and DSA.
Moreover, NFPA outperforms FPSA in iteration count for problem 1.

\section{Discussion}\label{sec4}

This paper introduced the Nonlinear Fokker-Planck Acceleration technique for steady-state, monoenergetic transport in homogeneous slab geometry.
To our knowledge, this is the first nonlinear HOLO method that accelerates \textit{all $L$ moments} of the angular flux.
Upon convergence, the LO and HO models are consistent; in other words, the (lower-order) modified Fokker-Planck equation \textit{preserves the same angular moments} of the flux obtained with the (higher-order) transport equation.

NFPA was tested on a homogeneous medium with an isotropic internal source with vacuum boundaries, and in a homogeneous medium with no internal source and an incoming beam boundary.
For both problems, three different scattering kernels were used.
The runtime and iterations of NFPA and FPSA were shown to be similar.
They both vastly outperformed DSA and GMRES for all cases by orders of magnitude.
However, NFPA has the feature of preserving the angular moments of the flux in both the HO and LO equations, which offers the advantage of integrating the LO model into multiphysics models. 

In the future, we intend to test NFPA capabilities for a variety of multiphysics problems and analyze its performance.
To apply NFPA to more realistic problems, it needs to be extended to include time and energy dependence. 
Additionally, the method needs to be adapted to address geometries with higher-order spatial dimensions.
Finally, for the NFPA method to become mathematically ``complete", a full convergence examination using Fourier analysis must be performed.
However, this is beyond the scope of this paper and must be left for future work.

\section*{Acknowledgements}

The authors acknowledge support under award number NRC-HQ-84-15-G-0024 from the Nuclear Regulatory Commission.
The statements, findings, conclusions, and recommendations are those of the authors and do not necessarily reflect the view of the U.S. Nuclear Regulatory Commission.

J.~K. Patel would like to thank Dr.~James Warsa for his wonderful transport class at UNM, as well as his synthetic acceleration codes.
The authors would also like to thank Dr.~Anil Prinja for discussions involving Fokker-Planck acceleration.

%\section*{References}

\bibliography{Kuczek_Patel_Vasques_arxiv}

\end{document}